# Modelling brain based on canonical ensemble with functional MRI: A thermodynamic exploration on neural system

Author:

Chenxi Zhou[1,2]     E-mail address: m201972604@hust.edu.cn

Bin Yang[1,2]     E-mail address: 836675486@qq.com

Wenliang Fan[3]     E-mail address: fwl@hust.edu.cn

Wei Li[1,2,*]     E-mail address: liwei0828@hust.edu.cn

For the Alzheimer's Disease Neuroimaging Initiative[**]

1. School of Artificial Intelligence and Automation, Huazhong University of Science and Technology, Wuhan, Hubei, 430074, China

2. Image Processing and Intelligent Control Key Laboratory of the Education Ministry of China, Wuhan, Hubei, 430074, China

3. Department of Radiology, Union Hospital, Tongji Medical College, Huazhong University of Science and Technology, Wuhan, Hubei, 430074, China

Corresponding author:

Wei Li

E-mail address: liwei0828@hust.edu.cn

Full postal address: School of Artificial Intelligence and Automation, Huazhong University of Science and Technology, Wuhan, Hubei, 430074, China.

[*] Corresponding author: Wei Li (E-mail address: liwei0828@hust.edu.cn)

Present/permanent address: School of Artificial Intelligence and Automation, Huazhong University of Science and Technology, Wuhan, Hubei, 430074, China.

[**] Data used in preparation of this article were obtained from the Alzheimer's Disease Neuroimaging Initiative (ADNI) database(adni.loni.usc.edu). As such, the investigators within the ADNI contributed to the design and implementation of ADNI and/or provided data but did not participate in analysis or writing of this report. A complete listing of ADNI investigators can be found at: http://adni.loni.usc.edu/wp-content/uploads/how_to_apply/ADNI_Acknowledgement_List.pdf

# Modelling brain based on canonical ensemble with functional MRI: A thermodynamic exploration on neural system

Chenxi Zhou[1,2], Bin Yang[1,2], Wenliang Fan[3], Wei Li[1,2,*], For the Alzheimer's Disease Neuroimaging Initiative[**]

1. School of Artificial Intelligence and Automation, Huazhong University of Science and Technology, Wuhan, Hubei, 430074, China
2. Image Processing and Intelligent Control Key Laboratory of the Education Ministry of China, Wuhan, Hubei, 430074, China
3. Department of Radiology, Union Hospital, Tongji Medical College, Huazhong University of Science and Technology, Wuhan, Hubei, 430074, China

Highlights:

(1) This study applied thermodynamics to the research of neural system in system level for the first time. Specifically, thermodynamic models of brain regions were built with extended canonical ensemble theory.

(2) The changed thermodynamic parameters, including higher internal energy, higher free energy and lower entropy in activated regions suggested that the neural systems also follow the laws of thermodynamics.

(3) By the detection of Alzheimer's disease, thermodynamic model was proved to be benefit from thermodynamic model, which meant the potential of thermodynamics in auxiliary diagnosis.

[*] Corresponding author: Wei Li (E-mail address: liwei0828@hust.edu.cn)
Present/permanent address: School of Artificial Intelligence and Automation, Huazhong University of Science and Technology, Wuhan, Hubei, 430074, China.
[**] Data used in preparation of this article were obtained from the Alzheimer's Disease Neuroimaging Initiative (ADNI) database(adni.loni.usc.edu). As such, the investigators within the ADNI contributed to the design and implementation of ADNI and/or provided data but did not participate in analysis or writing of this report. A complete listing of ADNI investigators can be found at: http://adni.loni.usc.edu/wp-content/uploads/how_to_apply/ADNI_Acknowledgement_List.pdf

Abstract：

*Objective*. Modelling is an important way to study the working mechanism of brain. While the characterization and understanding of brain are still inadequate. This study tried to build a model of brain from the perspective of thermodynamics at system level, which brought a new thinking to brain modelling.

*Approach*. Regarding brain regions as systems, voxels as particles, and intensity of signals as energy of particles, the thermodynamic model of brain was built based on canonical ensemble theory. Two pairs of activated regions and two pairs of inactivated brain regions were selected for comparison in this study, and the analysis on thermodynamic properties based on the model proposed were performed. In addition, the thermodynamic properties were also extracted as input features for the detection of Alzheimer's disease.

*Main results*. The experiment results verified the assumption that the brain also follows the thermodynamic laws. It demonstrated the feasibility and rationality of brain thermodynamic modelling method proposed, indicating that thermodynamic parameters could be applied to describe the state of neural system. Meanwhile, the brain thermodynamic model achieved much better accuracy in detection of Alzheimer's disease, suggesting the potential application of thermodynamic model in auxiliary diagnosis.

*Significance*. (1) Instead of applying some thermodynamic parameters to analyze neural system, a brain model at system level was proposed from perspective of thermodynamics for the first time in this study. (2) The study discovered that the neural system also follows the laws of thermodynamics, which leads to increased internal energy, increased free energy and decreased entropy when system is activated. (3) The detection of neural disease was demonstrated to be benefit from thermodynamic model, implying the immense potential of thermodynamics in auxiliary diagnosis.

# 1.Introduction

The human brain is the most complex organ with unlimited potential. In order to further understand the physiological principle and working mechanism of brain, scientists have tried to build models of brain from different perspectives and then we can observe and study system states and characteristics. Thereinto, dynamics is an important theory applied in modeling neural system, called neurodynamics. It attempts to build neural system models based on dynamics at different levels, from microscopic ion channel to macroscopic connection of cerebral cortex. Then system states and dynamic characteristics could be observed to explore the working mechanism of brain. Deco and Rolls (2005) studied the spiking mechanisms of synapses and neurons in biased competition with dynamic analysis. Based on neurodynamics, Le Van Quyen (2003) came up with a comprehensive framework to analyze the spatio-temporal characteristics of brain on large scale. Heller and Casey (2016) studied the temporal development of adolescents in emotion by building a corresponding model with neurodynamics, and explored the relationship between temporal dynamic changes of adolescent in emotion and mood/anxiety disorders. Amari and Maginu (1988) built autocorrelation associative memory model based on statistical neurodynamics to analyze non-equilibrium dynamical behaviors in recalling process. By simplifying KIII model, Harter and Kozma (2005) built KA model of aperiodic dynamics observed in cortical systems to understand intelligent behavior in biological agents.

The brain network is also significant in modeling neural system which is based on function integration and differentiation mechanism in neurophysiology. The brain network includes structural brain network and functional brain network. The structural brain network describes the anatomical links between units, for instance, the white matter fiber tracts (Bullmore and Sporns, 2009). While the functional brain network focuses on describing the interacts between units, for example, the functional correlation (Rowe, 2010), or the autocoherent oscillations of inhibitory neurons (Burns et al., 2010). The brain network modeling has also been applied to the research on pathologic mechanisms of various neural diseases. Xiang et al. (2013) explored the changes of shortest paths and clustering coefficients of functional brain network in Alzheimer's disease. Dubbelink et al. (2014) studied the changes of brain network topology in Parkinson's disease with magnetoencephalography. Jeong et al. (2014) studied the difference of functional brain network between epileptics and control group using the global mutual information and global efficiency of different bands with whole-brain magnetoencephalography. Brain network made people jump out of previous research focusing on one specific independent object, but tried to understand the physiological principle and working mechanism of neural system on different levels from perspective of the collaborative working mechanism of brain.

Actually, as brain is a multi-layered and multi-dimensional complex system, it's risky to study brain only from one single level or perspective. In fact, the characterization and understanding of brain are still inadequacy. We still need to try more new ways to model and analyze the brain.

Thermodynamics is one of the most important direction in physics. It focuses on the laws and physical properties of thermal motion and the evolution process of macroscopic matter systems consisted of microscopic particles. Scientists have tried to extend some concepts or theories of thermodynamics into other research fields for general physical state analysis at system level. In mechanical engineering, Zhang (2017) established a theory of non-equilibrium thermodynamics for studying the thermo-poro-mechanical modeling of saturated clays. Albertin et al. (2011) used computational thermodynamics to optimize the hardness and wear resistance of high chromium cast iron. Base on statistical thermodynamics, Kamiyama at al. (2016) proposed a "Hakoniwa" method to predict the properties of materials composed with different types of atoms, through calculating atomic energy.

In astronomy, Setare and Sheykhi (2010) studied the interaction between viscous dark energy and dark matter in RSII braneworld with thermodynamics. Whitehouse and Bate (2006) extended thermodynamics to the research on the collapse of molecular cloud cores, and proposed a three-dimensional algorithm to explore thermodynamic properties during our star formation. In biology, from a statistical thermodynamic point of view, Guo and Brooks (1997) proposed a method which made it possible to calculate all thermodynamic properties of a protein model with specific structure, and the free energy surface and compaction processes of the protein model proposed were characterized. Based on thermodynamics, Fischer et al. (1998) generated a distribution of mixed-canonical ensemble corresponding to different temperatures, and the crossing of energy barriers of RNA was observed by sampling from this distribution. These studies suggested that the thermodynamic theory is fundamental and universal to some extent. In other words, its ideas and methods not only go for the researches of traditional thermodynamic systems, but also be suitable for system modeling in other fields. Further, thermodynamics can often bring new analytical perspectives for our understanding and interpretation of systems in other subject areas.

In the study of brain, some researches have tried to analyze the brain by using parameters of informatics derived from thermodynamics. Zhang et al. (2015) used Tsallis entropy from discrete wavelet packet transform in the analysis of brain image, and achieved the identification of glioma, meningioma and other neural diseases. Wang, N. et al. (2018) explored the influence of occupational factors to brain complexity by calculating the brain entropy of seafarers and nonseafarers. Lebedev et al. (2016) evaluated the impact of lysergic acid diethylamide (LSD) on personality with the mixed-effects model based on changes in brain entropy and observed personality during the follow-up. Co-opting the concept of free energy in informatics, Friston (2010) proposed the free energy principle, holding that each biological self-organizing system in equilibrium will minimize the free energy to avoid the "surprise". Friston and Buzsaki (2016) tried to explain the optimization and control mechanism as well as function differentiation of the brain with the proposed free energy principle theoretically. Freeman and Vitiello (2006) defined the square of EEG amplitudes as the rate of free energy dissipation to measure how much work was done. These studies showed that it's feasible to observe or analyze brain with specific properties of informatics derived from thermodynamics, such as information entropy and free-energy in information theory. However, these studies are all limited to borrowing concepts from informatics for signal analysis.

Instead of starting from informatics as above researches, this article tried to construct a thermodynamic model of brain at system level based on the canonical ensemble theory. Based on the model proposed, we attempt to explore and characterize the working mechanism of neural system from a fully new point of view. We hope this would shed a new light on understanding of brain.

## 2. Materials and Methods

### 2.1 Data Acquisition

The data used in this study were obtained from the Alzheimer's Disease Neuroimaging Initiative (ADNI) database (http://adni.loni.usc.edu/). The ADNI was launched in 2003 as a public-private partnership, led by Principal Investigator Michael W.Weiner, MD. The primary goal of ADNI has been to test whether serial MRI, PET, other biological markers, and clinical and neuropsychological assessment can be combined to measure the progression of mild cognitive impairment (MCI) and early AD. For up-to-date information, see www.adni-info.org. 116 patients with Alzheimer's disease (Age: 74.6 ±7.5) and 174 healthy subjects (Age: 75.5 ±6.1) were recruited from ADNI as AD group and NC(Normal Control) group respectively. The demographic information of subjects is shown in Table 1. There is no significant difference between the two groups of data in terms of age or gender.

Table 1 The demographic information of subjects.

| Group | Number | Gender(male/female) | Age (year) |
|---|---|---|---|
| AD | 116 | 55/61 | 74.6 ±7.5 |
| NC | 174 | 97/77 | 75.5±6.1 |

## 2.2 Data preprocessing

The rs-fMRI images used in this paper were scanned by 3T Philips scanners. The specific operation parameters were as follows: slice thickness = 3.3mm, number of slices = 48, TR/TE = 3000ms/30ms, flip angle = 80°, imaging matrix = 64×64. Each series had 140 volumes. Data preprocessing adopts SPM8 and DPARSF (Yan and Zang, 2010). The process was as follows: First of all, the first ten frames were discarded for magnetization equilibrium. Slicing timing and realigning were carried out on time series. Subjects were excluded with head rotation exceeding 2°or head translation exceeding 2mm. The Montreal Neurological Institute (MNI) standard human brain template was used to normalize all the corrected image data. Then the images were smoothed by a 4×4×4 Gaussian kernel to decrease spatial noises. The global mean signal was removed to reduce the nonneuronal signal fluctuations. The whole brain was divided into 90 regions with the automated anatomical labeling (AAL) template.

## 2.3 Denoising with Point Process Method

The BOLD signal has been demonstrated to be the description of hemodynamic response to neural stimulation (Ogawa et al., 1990). Therefore, scientists tried to search extreme points of the BOLD signal corresponding to neural stimulation and applied these points to simulate a clear BOLD signal combining with standard hemodynamic equation (Buckner, 1998). Point process analysis holds that there are some significant feature points in the time series of a complex event, and the noise can be decreased to highlight the essential nature by extracting these feature points (Cox and Isham, 1980). So, this study used the point process method to remove noise in the Blood Oxygen Level-Dependent (BOLD) signal for high signal-to-noise ratio.

The BOLD signal was standardized firstly. In the time series, maximum points and minimum points were recorded and the subsequent extreme points were selected as a pair in chronological order. Then an impulse sequence was obtained by calculating the amplitude increment of every point pair per unit time. Afterwards, we convolved this impulse sequence with the hemodynamic response function proposed by Cohen (1997) and a clear BOLD signal was reconstructed.

**2.4. Statistical Thermodynamic Modeling Method**

Canonical ensemble is an important concept in statistical thermodynamics. In statistical thermodynamics, as the system has lots of different microscopic states under given macroscopic conditions, the measurement of a system is the average of multiple measurements of the system over time. In order to obtain the measurement at one point, we can measure lots of identical systems which are under the same macroscopic state at that point, and replace the time-average measurement with the average of a large number of identical systems at the same point. The set of a large number of identical systems which are under the same macroscopic state is called ensemble. Under this premise, the average measurement over time can be replaced by the ensemble average. For an isolated system in equilibrium, in which the energy, volume and number of particles have all been given, as the probability of possible microscopic states of the system follows the microcanonical distribution, we could analyze the system with the microcanonical ensemble theory. However, it's difficult for most systems to ensure its energy which is necessary for microcanonical ensemble analysis. So researchers generally study the closed system with given temperature, volume and number of particles, in which the probability of possible microscopic states of the system follows the canonical distribution. Then we could apply canonical ensemble to analyze thermodynamic properties of the system.

Some scientists extended the canonical ensemble to the research field of electricity, black hole, and RNA, etc. (Yip et al., 1996; Yang and Zhang, 2016; Fischer et al., 1998), and the achievements in these fields indicated that the canonical ensemble is universal, but not limited to traditional thermodynamic systems. Thus, we tried to introduce it to the study of neuroscience. Researches have shown that temperature of the human brain generally maintains at 36.9±0.4°C (Wang et al., 2014). Besides, the volumes and voxels of different brain functional regions are all certain for each subject in neuroimaging. If we consider the voxels as particles inside the brain region, the number of particles will be determined. Meanwhile, brain regions won't exchange voxels with outside, in other words, there is no "matter" exchange for the brain region. Then the brain region basically meets the requirements of closed system. So we believe that the brain region can be regarded as the thermodynamic system whose microscopic states conform to the canonical distribution. Then, we can build the brain thermodynamic model based on the theoretical framework of canonical ensemble.

Based on the analysis above, this article put forward a brain thermodynamic model. We regarded the brain functional region as a system constituted of nearly independent particles. Each voxel in neuroimaging corresponds to one nearly independent particle. The BOLD signal intensity of fMRI reflects the metabolic intensity: The higher BOLD signal means higher metabolic level, leading to the increase of oxygen consumption and the increase of energy consumption. Therefore, the amplitude of the reconstructed BOLD signal was supposed as the energy of corresponding voxel in this study, and the physiological activated intensity of each voxel can be evaluated by energy E. Because relevant computational data had been standardized, all calculated results were dimensionless.

$Z_i$ is the partition function of microscopic state of voxel at the $i$-th time point, which can be calculated as:

$$Z_i = \exp(-\beta E_i) \tag{1}$$

where $E_i$ is the energy of voxel at the $i$-th time point, and the specific expression of $\beta$ is shown as follows:

$$\beta = \frac{1}{kT} \tag{2}$$

where $k$ is Boltzmann constant in thermodynamics, which is equal to 1 to simplify the calculation in this article. The thermodynamic temperature $T$ is 310K, equal to 37 ℃, the theoretical temperature of the human brain. Since the whole brain could represent the environment of brain regions.

If we look on each time point of the BOLD time series to one voxel as a microscopic state, the partition function of the voxel can be calculated by summing all microscopic states, and the computational formula is described as:

$$Z_r = \sum_i Z_i = \sum_i \exp(-\beta E_i) \tag{3}$$

where $Z_r$ is the partition function of the voxel, $\sum_i Z_i$ is the sum of its microscopic states at all points in BOLD time series. Owing to the indistinguishability of identical principles and ignoring the weak interactions among particles, the partition function of brain region is the combination of that of all voxels inside. The partition function of brain region, $Z_{br}$, can be derived from $Z_r$:

$$Z_{br} = \frac{1}{N!} \prod_r Z_r \tag{4}$$

where $\prod_r Z_r$ is the quadrature of the partition function of all voxels in the brain region, and $N$ is the number of voxels.

Based on the theoretical framework of canonical ensemble, we constructed a brain thermodynamic model, regarding brain regions as systems, voxels as particles, the intensity of reconstructed BOLD signals as the energy of particles, and the points of fMRI time series as different microscopic states. We referred this model as BrainTDM. This model tried to describe the working mechanism of neural system from a thermodynamic point of view.

Based on the BrainTDM proposed in this paper, the internal energy, free energy and entropy of neural system could be defined to evaluate thermodynamic characters of brain regions.

The internal energy, $U$, of the brain region can be calculated as:

$$U = \sum_r U_r = \sum_r \frac{\sum_i \left[E_i \exp\left(-\frac{E_i}{kT}\right)\right]}{Z_r} \tag{5}$$

The free energy, $F$, of the brain region can be calculated as:

$$F = U - TS = -kT \ln Z_{br} = -\frac{kT \log_{10} Z_{br}}{\log_{10} e} \tag{6}$$

The entropy, $S$, of the brain region can be calculated as:

$$S = k \ln Z_{br} - k\beta \frac{\partial}{\partial \beta} \ln Z_{br} \tag{7}$$

$$U = -\frac{\partial}{\partial \beta} \ln Z_{br} \tag{8}$$

$$F = -kT \ln Z_{br} \tag{9}$$

$$S = k \ln Z_{br} - k\beta \frac{\partial}{\partial \beta} \ln Z_{br} = -\frac{F}{T} + \frac{U}{T} = \frac{U-F}{T} \tag{10}$$

For one specific brain region, internal energy represents the statistical average energy of all microscopic states in the system, free energy represents the energy of the brain region which could be used to do external work, and entropy represents chaos of the brain region after being affected by external environment. Base on thermodynamics, we supposed that when the brain region was activated and consumed external energy to do work, the energy contained in itself and the energy to do work would both rise, and its internal state would be more orderly compared with inactivated brain regions.

## 2.5 Experimental Design

In this experiment, the fMRI of brain was processed by SPM8 and divided into 90 brain regions with the AAL template. Point process method was applied to reconstruct clear BOLD signals as input for the following model. In modeling, brain regions were regarded as systems, voxels were regarded as particles, and the intensity of BOLD signals was regarded as the energy of particles. Then the brain thermodynamic model was built based on the ensemble theory. The related thermodynamic parameters were calculated, including the partition function $Z$, internal energy $U$, free energy $F$, and entropy $S$. More analyses were taken to explore potential applications of this model.

2.5.1 Experimental Paradigm I

In thermal physics, once thermodynamic system consumes external energy to do work, the entropy will decrease, and the internal energy as well as free energy will increase. We assumed that the system of brain also follows this rule. That is, when the brain regions are activated with neurons consuming external energy to fire synchronously and sequentially, the thermodynamic parameters of brain regions will change in the same way. In experiment paradigm I, we tried to evaluate the proposed modeling method by analyzing the thermodynamic parameter features of activated and inactivated cerebral regions. Since the default mode network (DMN) is most commonly shown to be activated when a person is not focused on the outside world and the brain is at wakeful rest (Raichle et al., 2001; Zhang and Raichle, 2010), we selected the medial prefrontal cortex (mPFC) which is a key brain region of the default network as the object of study. Meanwhile, considering the constant sound stimulation during fMRI scanning, we also selected the Heschl gyrus (HES), which are mainly located in the primary auditory cortex, as the object of study (Morosan, 2001). Furthermore, we selected the precentral gyrus (PreCG), which is also referred to as the primary motor region or primary motor cortex that belongs to task-positive network, and the olfactory cortex (OLF), which is a key component of the limbic system, as the control subjects (Zhang and Raichle, 2010). The brain regions selected for analysis were listed in Table 2.

Table 2 Selected brain regions.

| Activated brain regions | Inactivated brain regions |
|---|---|
| left/right medial prefrontal cortex | left/right Precentral gyrus |
| left/right Heschl gyrus | left/right Olfactory cortex |

By regarding brain regions as systems, voxels as particles and the intensity of reconstructed BOLD signals as the energy of particles, we constructed the brain thermodynamic models of regions selected. Then thermodynamic parameters of different brain regions were calculated based on models constructed, including the partition function, internal energy, free energy and entropy. By analyzing the differences between the thermodynamic parameters of the activated and inactivated brain regions, we tried to figure out if the neural system also follows the laws of thermodynamics.

2.5.2 experimental paradigm II

Experiment paradigm II hoped to explore the potential applications of the brain thermodynamic model proposed in computer-aided diagnosis (CAD) or other realistic scenarios. For this purpose, we tried to recognize the subjects with Alzheimer's disease, taking parameters derived from the brain thermodynamic model as input features. In the meantime, we did the same classification task with parameters based on traditional brain network model as a comparison. Specific experiments were as follows:

(A) Thermodynamic Parameters - Kendall (TP-Kendall): This experiment constructed the brain thermodynamic model by regarding brain regions as systems and voxels as particles. Then four thermodynamic parameters could be obtained based on this model, namely, the partition function, internal energy, free energy and entropy. Since 90 brain regions were divided from cerebral cortex in AAL template, 360 thermodynamic parameters could be derived as alternative features. And 72 parameters with the largest Kendall tau rank correlation coefficients were chosen as input features of classifiers.

(B) Thermodynamic Parameters - Expert (TP-Expert): The modeling process of this experiment was the same as that of TP-Kendall, except that we merely chose the thermodynamic parameters from 18 brain regions highly associated with Alzheimer's disease based on expertise as features for classification. Then we also got 72 features, as each region has four thermodynamic parameters.

(C) Brain Network - Kendall (BN-Kendall): This experiment built the functional brain network model traditionally using Pearson correlation coefficient to obtain link intensities between brain regions. And then 72 link strengths between brain regions with the largest Kendall tau rank correlation coefficient were chosen as the input features of classifiers.

All three experiments selected 72 features for AD detection with KNN classifier.

## 3. Results

Based on the brain thermodynamic model proposed in this paper, we built models of neural system in experimental paradigm I and II and the results were shown below. In order to avoid the influence of accidental factors, ten-fold cross-validation strategy was applied in all classification tasks.

3.1 experimental paradigm I

Based on the modeling method proposed in this paper, we built the brain thermodynamic model of selected brain regions and obtained the energy of activated and inactivated brain regions for each subject. In NC group, we calculated the average energy of each brain region separately and then get the corresponding time series trajectory, as shown in Fig. 1. It was shown that the average energy of activated brain regions, mPFC/HES, were significantly higher than that of inactivated brain regions, PreCG/OLF.

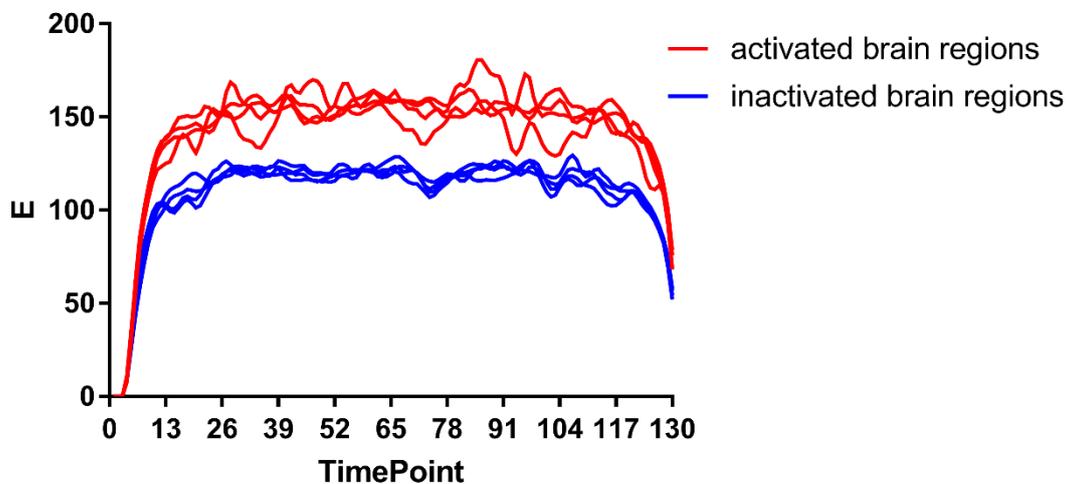

Fig. 1 Average energy trajectories of activated and inactivated brain regions. The red curves

represented the average energy $E$ of activated brain regions mPFC.L/mPFC.R/HES.L/HES.R, and the blue curves represented the average energy $E$ of inactivated brain regions PreCG.L/PreCG.R/OLF.L/OLF.R.

Based on the model built, the thermodynamic parameters of each brain region could be calculated, including partition function, internal energy, free energy and entropy. As the value of partition function was too large, we took the logarithm in calculation of this parameter. We calculated the mean of the thermodynamic parameters of different brain regions separately over NC group, and the results were shown in Fig. 2. There was no significant difference between activated brain regions, mPFC and HES, in these thermodynamic parameters. And we observed the same phenomenon between inactivated brain regions, PreCG and OLF. However, there was significant difference between activated and inactivated brain regions. The mean value of partition function and entropy in activated brain regions were lower than that in inactivated brain regions significantly, and the internal energy and free energy were the exact opposite

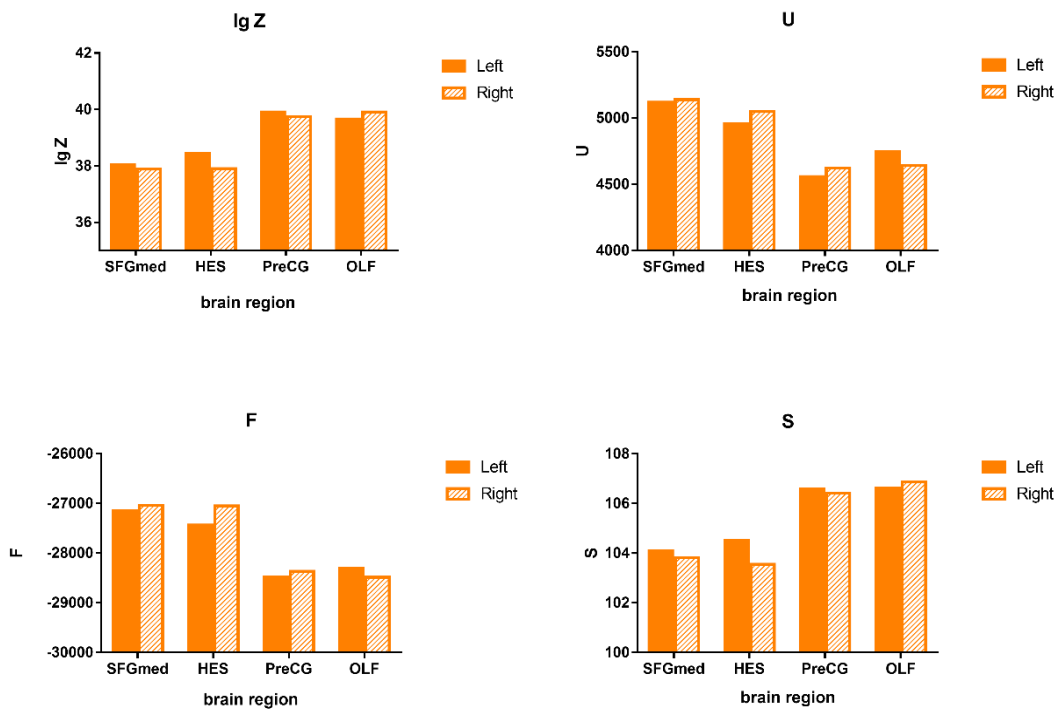

Fig. 2 Thermodynamic parameters of activated and inactivated brain regions. Left and Right represent left brain regions and right brain regions respectively.

T-test was applied to thermodynamic parameters of different brain regions and the results were shown in Table 3-6. It was shown that there were significant differences between activated and inactivated brain regions in all four thermodynamic parameters, partition function, internal energy, free energy and entropy($P<10^{-6}$). While there was no significant difference almost in all four thermodynamic parameters among four activated brain regions, and the same result was observed among four inactivated brain regions.

Table 3 T-test for lgZ.

| Brain Region | mPFC.L | mPFC.R | HES.L | HES.R | PreCG.L | PreCG.R | OLF.L | OLF.R |
|---|---|---|---|---|---|---|---|---|
| mPFC.L | 1 | | | | | | | |
| mPFC.R | 0.25 | 1 | | | | | | |
| HES.L | 0.05 | 0.01 | 1 | | | | | |
| HES.R | 0.58 | 0.96 | 0.07 | 1 | | | | |
| PreCG.L | 0.00 | 0.00 | 0.00 | 0.00 | 1 | | | |
| PreCG.R | 0.00 | 0.00 | 0.00 | 0.00 | 0.13 | 1 | | |
| OLF.L | 0.00 | 0.00 | 0.00 | 0.00 | 0.09 | 0.54 | 1 | |
| OLF.R | 0.00 | 0.00 | 0.00 | 0.00 | 0.96 | 0.22 | 0.13 | 1 |

There is significant difference when P-value is less than 0.05.

Table 4 T-test for U.

| Brain Region | mPFC.L | mPFC.R | HES.L | HES.R | PreCG.L | PreCG.R | OLF.L | OLF.R |
|---|---|---|---|---|---|---|---|---|
| mPFC.L | 1 | | | | | | | |
| mPFC.R | 0.67 | 1 | | | | | | |
| HES.L | 0.01 | 0.01 | 1 | | | | | |
| HES.R | 0.26 | 0.15 | 0.23 | 1 | | | | |
| PreCG.L | 0.00 | 0.00 | 0.00 | 0.00 | 1 | | | |
| PreCG.R | 0.00 | 0.00 | 0.00 | 0.00 | 0.15 | 1 | | |
| OLF.L | 0.00 | 0.00 | 0.00 | 0.00 | 0.00 | 0.03 | 1 | |
| OLF.R | 0.00 | 0.00 | 0.00 | 0.00 | 0.12 | 0.70 | 0.10 | 1 |

There is significant difference when P-value is less than 0.05.

Table 5 T-test for F .

| Brain Region | mPFC.L | mPFC.R | HES.L | HES.R | PreCG.L | PreCG.R | OLF.L | OLF.R |
|---|---|---|---|---|---|---|---|---|
| mPFC.L | 1 | | | | | | | |

| Brain Region | mPFC.L | mPFC.R | HES.L | HES.R | PreCG.L | PreCG.R | OLF.L | OLF.R |
|---|---|---|---|---|---|---|---|---|
| mPFC.R | 0.25 | 1 | | | | | | |
| HES.L | 0.05 | 0.01 | 1 | | | | | |
| HES.R | 50.58 | 0.96 | 0.07 | 1 | | | | |
| PreCG.L | 0.00 | 0.00 | 0.00 | 0.00 | 1 | | | |
| PreCG.R | 0.00 | 0.00 | 0.00 | 0.00 | 0.13 | 1 | | |
| OLF.L | 0.00 | 0.00 | 0.00 | 0.00 | 0.09 | 0.54 | 1 | |
| OLF.R | 0.00 | 0.00 | 0.00 | 0.00 | 0.96 | 0.22 | 0.13 | 1 |

There is significant difference when P-value is less than 0.05.

Table 6 T-test for S.

| Brain Region | mPFC.L | mPFC.R | HES.L | HES.R | PreCG.L | PreCG.R | OLF.L | OLF.R |
|---|---|---|---|---|---|---|---|---|
| mPFC.L | 1 | | | | | | | |
| mPFC.R | 0.11 | 1 | | | | | | |
| HES.L | 0.18 | 0.03 | 1 | | | | | |
| HES.R | 0.19 | 0.53 | 0.05 | 1 | | | | |
| PreCG.L | 0.00 | 0.00 | 0.00 | 0.00 | 1 | | | |
| PreCG.R | 0.00 | 0.00 | 0.00 | 0.00 | 0.15 | 1 | | |
| OLF.L | 0.00 | 0.00 | 0.00 | 0.00 | 0.82 | 0.25 | 1 | |
| OLF.R | 0.00 | 0.00 | 0.00 | 0.00 | 0.07 | 0.01 | 0.22 | 1 |

There is significant difference when P-value is less than 0.05.

Furthermore, we tried to classify brain regions into activated and inactivated ones using thermodynamic parameters as input features with KNN classifier. Three kinds of distance measurements were respectively taken in KNN, including correlation distance, cosine distance and Euclidean distance. The results were shown in Table 7. The classification accuracy fluctuated around 90%, reaching the highest 91.1% (Cosine distance, K=3).

Table 7 The accuracy of classification for activated/inactivated brain regions (%).

| K \ Distance | Correlation | Cosine | Euclidean |
|---|---|---|---|
| 1 | 87.4 | 88.6 | 88.7 |
| 2 | 86.6 | 88.8 | 87.9 |
| 3 | 89.8 | 91.1 | 90.4 |

## 3.2 Experiment Paradigm Ⅱ

In experiment paradigm Ⅱ, we tried to apply the brain thermodynamic model proposed to the CAD of Alzheimer's disease. We designed three experiments, namely TP-Kendall, TP-Expert and BN-Kendall for comparison. These experiments also used KNN as classifier based on correlation distance, cosine distance and Euclidean distance respectively. Results were shown in Table 8. The accuracy of TP-Kendall ranged from 77.0% to 83.7%, with an average of 81.6% and a maximum of 83.7%; the accuracy of TP-Expert ranged from 68.3% to 73.7%, with an average of 70.6% and a maximum of 73.7%; and the accuracy of BN-Kendall ranged from 56.9% to 62.1%, with an average of 59.1% and a maximum of 62.1%. It was to say, no matter Kendall coefficient or expertise was used for feature selection, the recognition accuracy of the proposed brain thermodynamic model was significantly higher than that of the brain network model. In particular, with the same feature selection method, the accuracy of Alzheimer's disease recognition based on thermodynamic model (TP-Kendal) was 22.5% higher than that of the brain network model (BN-Kendall) on average.

Table 8 The accuracy of classification for AD/NC (%).

| Feature selection / Distance | TP-Kendall | | | TP-Expert | | | BN-Kendall | | |
|---|---|---|---|---|---|---|---|---|---|
| | K=1 | K=2 | K=3 | K=1 | K=2 | K=3 | K=1 | K=2 | K=3 |
| Correlation | 83.3 | 81.0 | 83.3 | 72.3 | 71.7 | 73.7 | 62.1 | 58.0 | 57.4 |
| Cosine | 83.7 | 81.4 | 83.6 | 70.1 | 69.9 | 71.0 | 61.1 | 58.9 | 60.8 |
| Euclidean | 80.2 | 77.0 | 80.8 | 70.5 | 67.5 | 68.3 | 60.2 | 56.9 | 56.9 |

In this study, Both TP-Kendall and BN-Kendall experiments used Kendall for feature selection. Based on these two experiments, we analyzed the impact of different input features on Alzheimer's disease recognition, also using KNN classifier as above. The results were shown in Fig. 3. Regardless of the number of input features, the accuracy of Alzheimer's disease recognition with thermodynamic parameters from the model proposed in this paper was always much higher than that with link strengths from brain network model, 17.2% higher on average. Furthermore, the accuracy of Alzheimer's disease recognition based on the brain thermodynamic model reached 86.4% with only 360 input features in correlation distance. While that based on the brain network model peaked only at 67.6% with 4005 input features in correlation distance.

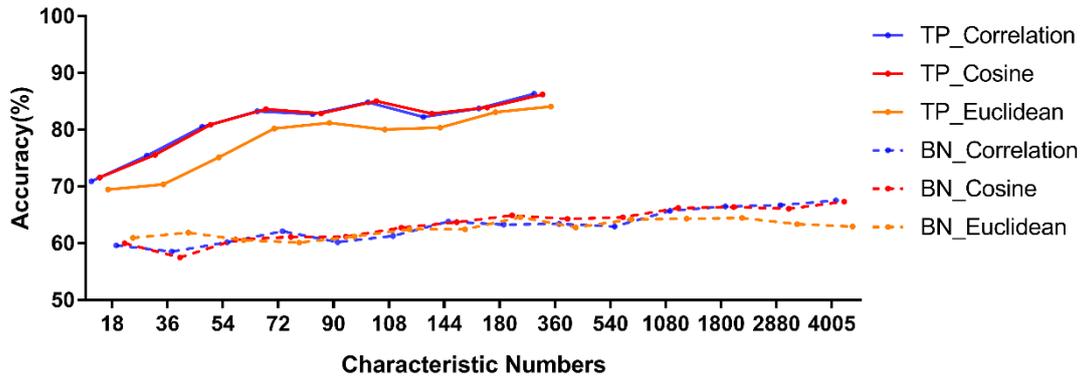

Fig. 3 The impact of feature number on accuracy of AD/NC classification.

## 4. Discussion

The brain is the most complex system in the known world. Brain modeling is an effective way to explore the work mechanism of brain, and is one of the hottest research topics all the time. Some scientists attempted to model the brain based on the neurodynamic principles and methods, from microscopic ion channel layer to macroscopic neural network layer (Gerstner et al., 2014). Most initial researches focused on the basic working mechanism of neurons or neural connectivity (Deco and Rolls, 2005; Le Van Quyen, 2003; Jiao and Wang, 2005). Later, scientists began to use neurodynamic method to explain some complicated brain functions, such as emotion, language acquisition and language comprehension (Heller and Casey, 2016; Partanen et al. 2017; Armeni et al., 2019), and then developed this method to analyze the pathological mechanism of epilepsy, Alzheimer's disease and other neural system diseases (Van Quyen et al., 2003; Jeong, 2002). In addition, from another perspective, brain differentiation and integration, scientists proposed the modeling method of brain network. Early researches mainly concentrated on exploring how to construct brain network from the neuron layer to the functional cortex layer. The structure equation, causality, correlation and consistency were applied to the definition of structure or function connectivity of brain (Bullmore and Sporns, 2009; He et al., 2009). Moreover, scientists used brain network to explore the impact of brain lesions on coupling brain regions (Filippi et al., 2013) and the pathological mechanism of neural system diseases, such as brachial plexus injury and Alzheimer's disease (Wang, W. W. et al., 2018; Yao et al., 2010). Some studies tried to extract characteristics of brain network for the identification, prediction and prognosis of diseases and got really good results (Wee et al., 2012; Jie et al., 2014; Aerts et al., 2020).

These modeling methods of brain have got many meaningful results. However, brain is a typical multi-dimensional complex system, and the understanding of the brain is far inadequacy till now, especially from the point of system view.

Thermodynamics, a major branch of physics, mainly studies the thermal properties of the object from the perspective of energy conversion on the macro level. Based on observed phenomenon in experiment, thermodynamics applies mathematical modeling methods to draw relevant conclusions by logical deduction. Thus, it belongs to phenomenological theory, indicating that the conclusions drawn from that are highly reliable and universal. Therefore, people always tried to extend and apply related mature theories and concepts of thermodynamics to system modeling and analysis of other research fields, such as mechanical engineering (Zhang, 2007; Albertin et al., 2011; Kamiyama et al., 2016), astronomy (Setare and Sheykhi, 2010; Whitehouse and Bate, 2006), biology (Guo and Brooks, 1997; Fischer et al., 1998) and other non-classical physical fields, so as to realize generalized physical state analysis of objects at the system level in different areas. In this study, we deemed that the material foundation which the brain realizes various functions based on is energy conversion. Thus, we assumed that the brain also follows the related macroscopic laws of energy conversion in thermodynamics. When neurons are activated, brain will also consume the external energy input to do work as thermodynamic systems, which leads to increased internal energy, increased free energy, and decreased entropy of system. Based on the above assumption, we proposed a thermodynamic model of brain for the first time. Taking brain regions as systems, voxels as particles and the intensity of BOLD signals as the energy of particles, this method built the brain thermodynamic model (BrainTDM) and tried to explain the work mechanism of brain based on the canonical ensemble theory from the perspective of energy conversion in thermodynamics.

In experimental paradigm I, we selected two pairs of activated brain regions and two pairs of inactivated brain regions at resting state as objects. Then the BrainTDMs of regions were built and thermodynamic parameters were calculated respectively. The experiment results showed that the internal energy and free energy of activated brain regions (mPFC and HES) were all much higher than those parameters of inactivated regions (PreCG and OLF), while it was opposite for the partition function and entropy. It exactly validated the assumption that brain also obeys the laws of thermodynamics at the system level. When activated, the neurons of the specific region burn energy to generate electrical impulses for information transmission, which is doing work just as those thermodynamic systems, leading to the corresponding thermodynamic parameter change of regions: internal energy and free energy increasing, and entropy decreasing. In other words, the brain also follows the laws of energy conversion in thermodynamic systems on macroscopic level. It is the most important discovery in this study. The results demonstrated that the brain thermodynamic model proposed in this paper is workable, and constructing brain model based on canonical ensemble theory of thermodynamics is feasible.

We also observed that the energy of activated brain regions was significantly higher than that of inactivated ones. It is in keeping with the common cognition that the activation levels of mPFC and HES were significantly higher than that of PreCG and OLF at resting state. For activated regions, the enhancement of DMN activation at resting state has been demonstrated in many papers, also consistent with the observation of fMRI imaging in experiment. The activation of HES, auditory related, was due to the continuous noises from nuclear magnetic resonance equipment in experiment. For inactivated regions, PreCG and OLF are related with the control of movement and olfaction respectively, and were considered to be inactivated due to resting state. In addition, we tried to classify the brain regions into the activated ones and inactivated ones with KNN classifier using thermodynamic parameters as input features. The experimental results showed that the classification accuracy kept around 90% and reached the highest 91.1% in three kinds of distance measurements, correlation distance, cosine distance and Euclidean distance, which indicated that thermodynamic parameters could actually reflect the differences of energy conversion state between activated and inactivated brain regions. In thermodynamics, the internal energy, free energy, entropy and other thermodynamic parameters were regarded as the common state functions describing the state of system. This study implied that thermodynamic parameters obtained from the model proposed could also be used as state functions of brain regions to characterize the activated state of brain regions.

The experimental paradigm II tried to apply the brain thermodynamic model proposed to discriminate diseases of neural system. In this study, using thermodynamic parameters from the model proposed, the detection of Alzheimer's disease could achieve far better results than that using link strengths from traditional brain network model (above 17.2% higher accuracy on average), and reached 86.4% with only 360 input features. It demonstrated that the brain thermodynamic parameters really contain some pathological information of neural diseases in nature. And this information must be essential and critical in describing the change of brain with diseases, as we could get 71.6% accuracy of AD detection with only 18 thermodynamic parameters as input features in cosine distance. While the brain network model could only get 60.0% accuracy using the same number of features under the same conditions. The above results indicated the brain thermodynamic model proposed in this paper could not only explain the basic working mechanism of neural system from thermodynamics, but also have the potential to be applied to recognition and prediction of neural system diseases. On the other hand, the experimental results also implied that the detection of neural system diseases may be benefit from the laws of energy conversion in neural system.

## 5. Conclusion

Scientists have always tried to extend the thermodynamic theories and relative concepts to system modelling and system analysis in mechanical engineering, astronomy, biology and other fields. Taking brain regions as systems, voxels as particles, and the intensity of BOLD signals as the energy of particles, this study mapped neural system to thermodynamic system. Based on canonical ensemble theory, the BrainTDM was built to explore the work mechanism of brain. The experiment results demonstrated the feasibility and rationality of modelling the neural system from the perspective of thermodynamics, and on the other hand, verified the hypothesis that the brain also follows the laws of thermodynamics. In addition, the study also indicated the positive impacts of the laws of energy conversion on the detection of neural system diseases, implying the potentiality of the model to be applied to the auxiliary diagnosis.

**Acknowledgement:**

This work was supported by the National Natural Science Foundation of China (61473131). Data collection and sharing for this project was funded by the Alzheimer's Disease Neuroimaging Initiative (ADNI) (National Institutes of Health Grant U01 AG024904) and DOD ADNI (Department of Defense award number W81XWH-12-2-0012). ADNI is funded by the National Institute on Aging, the National Institute of Biomedical Imaging and Bioengineering, and through generous contributions from the following: AbbVie, Alzheimer's Association; Alzheimer's Drug Discovery Foundation; Araclon Biotech; BioClinica, Inc.; Biogen; Bristol-Myers Squibb Company; CereSpir, Inc.; Cogstate; Eisai Inc.; Elan Pharmaceuticals, Inc.; Eli Lilly and Company; EuroImmun; F. Hoffmann-La Roche Ltd and its affiliated company Genentech, Inc.; Fujirebio; GE Healthcare; IXICO Ltd.; Janssen Alzheimer Immunotherapy Research & Development, LLC.; Johnson & Johnson Pharmaceutical Research & Development LLC.; Lumosity; Lundbeck; Merck & Co., Inc.; Meso Scale Diagnostics, LLC.; NeuroRx Research; Neurotrack Technologies; Novartis Pharmaceuticals Corporation; Pfizer Inc.; Piramal Imaging; Servier; Takeda Pharmaceutical Company; and Transition Therapeutics. The Canadian Institutes of Health Research is providing funds to support ADNI clinical sites in Canada. Private sector contributions are facilitated by the Foundation for the National Institutes of Health (www.fnih.org). The grantee organization is the Northern California Institute for Research and Education, and the study is coordinated by the Alzheimer's Therapeutic Research Institute at the University of Southern California. ADNI data are disseminated by the Laboratory for Neuro Imaging at the University of Southern California.

**Data and code availability statement:**

Data from the Alzheimer's Disease Neuroimaging Initiative (ADNI) which was used and analyzed in this study can be download at ADNI online repositories: http://adni.loni.usc.edu/. The MATLAB code for the processing, modelling and analysis in this study is available from the corresponding author upon request.

**Credit author statement:**

**Chenxi Zhou**: Methodology, Software, Formal analysis, Writing. **Bin Yang:** Methodology, Software. **Wenliang Fan:** Formal Analysis, Investigation. **Wei Li**: Conceptualization, Methodology, Formal analysis, Writing, Funding acquisition.

**Declarations of interest:**

None.

# Reference

Aerts, H., Schirner, M., Dhollander, T., Jeurissen, B., Achten, E., Van Roost, D., Ritter, P., Marinazzo, D., 2020. Modeling brain dynamics after tumor resection using The Virtual Brain. Neuroimage. 213, 116738. https://doi.org/10.1016/j.neuroimage.2020.116738.

Albertin, E., Beneduce, F., Matsumoto, M., Teixeira, I., 2011. Optimizing heat treatment and wear resistance of high chromium cast irons using computational thermodynamics. Wear. 271(9-10), 1813-1818. https://doi.org/10.1016/j.wear.2011.01.079.

Amari S. I., Maginu K., 1988. Statistical neurodynamics of associative memory. Neural Networks. 1(1):63-73. https://doi.org/10.1016/0893-6080(88)90022-6.

Armeni, K., Willems, R. M., van den Bosch, A., Schoffelen, J. M., 2019. Frequency-specific brain dynamics related to prediction during language comprehension. Neuroimage. 198, 283-295. https://doi.org/10.1016/j.neuroimage.2019.04.083.

Buckner, R., 1998. Event-related fMRI and the hemodynamic response. Human Brain Mapping. 6(5-6), 373-377. https://doi.org/10.1002/(SICI)1097-0193(1998)6:5/6<373::AID-HBM8>3.0.CO;2-P.

Bullmore, E., Sporns, O., 2009. Complex brain networks: graph theoretical analysis of structural and functional systems. Nature Reviews Neuroscience. 10(3), 186-198. https://doi.org/10.1038/nrn2575.

Burns, S. P., Xing, D., Shelley, M. J., Shapley, R. M., 2010. Searching for autocoherence in the cortical network with a time-frequency analysis of the local field potential. J Neurosci. 30(11), 4033-4047. https://doi.org/10.1523/JNEUROSCI.5319-09.2010.

Cohen, M., 1997. Parametric analysis of fMRI data using linear systems methods. Neuroimage. 6(2), 93-103. https://doi.org/10.1006/nimg.1997.0278.

Cox, D.R., Isham, V., 1980. Point Processes, 1st ed. CRC Press, Boca Raton. https://doi.org/10.1201/9780203743034.

Deco, G., Rolls, E., 2005. Neurodynamics of biased competition and cooperation for attention: A model with spiking neurons. Journal of Neurophysiology. 94(1), 295-313. https://doi.org/10.1152/jn.01095.2004.

Dubbelink, K., Hillebrand, A., Stoffers, D., Deijen, J., Twisk, J., Stam, C., Berendse, H., 2014.

Disrupted brain network topology in Parkinson's disease: a longitudinal magnetoencephalography study. Brain. 137, 197-207. https://doi.org/10.1093/brain/awt316.

Filippi, M., Agosta, F., Scola, E., Canu, E., Magnani, G., Marcone, A., Valsasina, P., Caso, F., Copetti, M., Comi, G., Cappa, S. F., Falini, A., 2013. Functional network connectivity in the behavioral variant of frontotemporal dementia. Cortex. 49(9), 2389-2401. https://doi.org/10.1016/j.cortex.2012.09.017.

Fischer, A., Cordes, F., Schutte, C., 1998. Hybrid Monte Carlo with adaptive temperature in mixed-canonical ensemble: Efficient conformational analysis of RNA. Journal of Computational Chemistry. 19(15), 1689-1697. https://doi.org/10.1002/(SICI)1096-987X(19981130)19:15<1689::AID-JCC2>3.3.CO;2-R

Freeman, W., Vitiello, G., 2006. Nonlinear brain dynamics as macroscopic manifestation of underlying many-body field dynamics. Physics of Life Reviews. 3(2), 93-118. https://doi.org/10.1016/j.plrev.2006.02.001.

Friston, K., 2010. The free-energy principle: a unified brain theory?. Nature Reviews Neuroscience. 11(2), 127-138. https://doi.org/10.1038/nrn2787.

Friston, K., Buzsaki, G., 2016. The Functional Anatomy of Time: What and When in the Brain. Trends in Cognitive Sciences. 20(7), 500-511. https://doi.org/10.1016/j.tics.2016.05.001.

Gerstner W., Kistler W.M., Naud R., Paninski L., 2014.Neuronal Dynamics: From Single Neurons to Networks and Models of Cognition. Cambridge University Press, New York. https://doi.org/10.1017/CBO9781107447615.

Guo, Z., Brooks, C., 1997. Thermodynamics of protein folding: A statistical mechanical study of a small all-beta protein. Biopolymers. 42(7), 745-757. https://doi.org/10.1002/(SICI)1097-0282(199712)42:7<745::AID-BIP1>3.0.CO;2-T.

Harter, D., Kozma, R., 2005. Chaotic neurodynamics for autonomous agents. IEEE Trans Neural Netw. 16(3), 565-579. https://doi.org/10.1109/TNN.2005.845086.

He, Y., Chen, Z., Gong, G., Evans, A., 2009. Neuronal Networks in Alzheimer's Disease. Neuroscientist. 15(4), 333-350. https://doi.org/10.1177/1073858409334423.

Heller, A. S., Casey, B. J., 2016. The neurodynamics of emotion: delineating typical and atypical emotional processes during adolescence. Dev Sci. 19(1), 3-18.

https://doi.org/10.1111/desc.12373.

Jeong, J., 2002. Nonlinear dynamics of EEG in Alzheimer's disease. Drug Development Research. 56(2), 57-66. https://doi.org/10.1002/ddr.10061.

Jeong, W., Jin, S., Kim, M., Kim, J., Chung, C., 2014. Abnormal functional brain network in epilepsy patients with focal cortical dysplasia. Epilepsy Research. 108(9), 1618-1626. https://doi.org/10.1016/j.eplepsyres.2014.09.006

Jiao, X., Wang, R., 2005. Nonlinear dynamic model and neural coding of neuronal network with the variable coupling strength in the presence of external stimuli. Applied Physics Letters. 87(8), e083901. https://doi.org/10.1063/1.1957120.

Jie, B., Zhang, D., Gao, W., Wang, Q., Wee, C. Y., Shen, D., 2014. Integration of network topological and connectivity properties for neuroimaging classification. IEEE Trans Biomed Eng. 61(2), 576-589. https://doi.org/10.1109/TBME.2013.2284195.

Kamiyama, E., Matsutani, R., Suwa, R., Vanhellemont, J., Sueoka, K., 2016. The Hakoniwa method, an approach to predict material properties based on statistical thermodynamics and ab initio calculations. Materials Science in Semiconductor Processing. 43, 209-213. https://doi.org/10.1016/j.mssp.2015.12.023.

Le Van Quyen, M., 2003. Disentangling the dynamic core: a research program for a neurodynamics at the large-scale. Biological Research. 36(1), 67-88. https://doi.org/10.4067/S0716-97602003000100006.

Lebedev, A. V., Kaelen, M., Lövdén, M., Nilsson, J., Feilding, A., Nutt, D. J., Carhart-Harris, R. L., 2016. LSD-induced entropic brain activity predicts subsequent personality change. Hum Brain Mapp. 37(9), 3203-3213. https://doi.org/10.1002/hbm.23234.

Morosan P., Rademacher J., Schleicher A., Amunts K., Schormann T., Zilles K., 2001. Human primary auditory cortex: cytoarchitectonic subdivisions and mapping into a spatial reference system. Neuroimage. 13, 684–701. https://doi.org/10.1006/nimg.2000.0715.

Ogawa, S., Lee, T. M., Nayak, A. S., Glynn, P., 1990. Oxygenation-sensitive contrast in magnetic resonance image of rodent brain at high magnetic fields. Magn Reson Med. 14(1), 68-78. https://doi.org/10.1002/mrm.1910140108.

Partanen, E., Leminen, A., de Paoli, S., Bundgaard, A., Kingo, O. S., Krøjgaard, P., Shtyrov, Y.,

2017. Flexible, rapid and automatic neocortical word form acquisition mechanism in children as revealed by neuromagnetic brain response dynamics. Neuroimage. 155, 450-459. https://doi.org/10.1016/j.neuroimage.2017.03.066

Raichle, M.E., MacLeod, A.M., Snyder, A.Z., Powers, W.J., Gusnard, D.A., Shulman, G.L., 2001. A default mode of brain function. PNAS. 98(2), 676-682. https://doi.org/10.1073/pnas.98.2.676.

Rowe, J. B., 2010. Connectivity Analysis is Essential to Understand Neurological Disorders. Front Syst Neurosci. 4, e00144. https://doi.org/10.3389/fnsys.2010.00144.

Setare, M., Sheykhi, A., 2010. THERMODYNAMICS OF VISCOUS DARK ENERGY IN AN RSII BRANEWORLD. International Journal of Modern Physics D. 19(2), 171-181. https://doi.org/10.1142/S0218271810016361.

Van Quyen, M., Navarro, V., Martinerie, J., Baulac, M., Varela, F., 2003. Toward a neurodynamical understanding of ictogenesis. Epilepsia. 44, 30-43. https://doi.org/10.1111/j.0013-9580.2003.12007.x.

Wang, H., Wang, B., Normoyle, K., Jackson, K., Spitler, K., Sharrock, M., Miller, C., Best, C., Llano, D., Du, R., 2014. Brain temperature and its fundamental properties: a review for clinical neuroscientists. Frontiers in Neuroscience, 8, e00307. https://doi.org/10.3389/fnins.2014.00307.

Wang, N., Wu, H., Xu, M., Yang, Y., Chang, C., Zeng, W., Yan, H., 2018. Occupational functional plasticity revealed by brain entropy: A resting-state fMRI study of seafarers. Hum Brain Mapp. 39(7), 2997-3004. https://doi.org/10.1002/hbm.24055.

Wang, W. W., Lu, Y. C., Tang, W. J., Zhang, J. H., Sun, H. P., Feng, X. Y., Liu, H. Q., 2018. Small-worldness of brain networks after brachial plexus injury: A resting-state functional magnetic resonance imaging study. Neural Regen Res. 13(6), 1061-1065. https://doi.org/10.4103/1673-5374.233450

Wee, C. Y., Yap, P. T., Denny, K., Browndyke, J. N., Potter, G. G., Welsh-Bohmer, K. A., Wang, L., Shen, D., 2012. Resting-state multi-spectrum functional connectivity networks for identification of MCI patients. PLoS One. 7(5), e37828. https://doi.org/10.1371/journal.pone.0037828

Whitehouse, S., Bate, M., 2006. The thermodynamics of collapsing molecular cloud cores using smoothed particle hydrodynamics with radiative transfer. Monthly Notices of the Royal Astronomical Society. 367(1), 32-38. https://doi.org/10.1111/j.1365-2966.2005.09950.x.

Xiang, J., Guo, H., Cao, R., Liang, H., Chen, J., 2013. An abnormal resting-state functional brain network indicates progression towards Alzheimer's disease. Neural Regen Res. 8(30), 2789-2799. https://doi.org/10.3969/j.issn.1673-5374.2013.30.001.

Yan, C. G., Zang Y. F., 2010. DPARSF: A MATLAB Toolbox for "Pipeline" Data Analysis of Resting-State fMRI. Front Syst Neurosci. 4, 13. https://doi.org/10.3389/fnsys.2010.00013.

Yang, J., He, T., Zhang, J., 2016. Simple Harmonic Oscillator Canonical Ensemble Model for Tunneling Radiation of Black Hole. Entropy, 18(11), e18110415. https://doi.org/10.3390/e18110415.

Yao, Z., Zhang, Y., Lin, L., Zhou, Y., Xu, C., Jiang, T., Initiative, A. s. D. N., 2010. Abnormal cortical networks in mild cognitive impairment and Alzheimer's disease. PLoS Comput Biol, 6(11), e1001006. https://doi.org/10.1371/journal.pcbi.1001006.

Yip, M. K., Zheng, J. R., Cheung, H. F., 1996. Persistent current of one-dimensional perfect rings under the canonical ensemble. Phys Rev B Condens Matter. 53(3), 1006-1009. https://doi.org/10.1103/physrevb.53.1006

Zhang, D., Raichle, M.E., 2010. Disease and the brain's dark energy. Nat Rev Neurol. 6(1), 15-28. https://doi.org/10.1038/nrneurol.2009.198.

Zhang, Y., Dong, Z., Wang, S., Ji, G., Yang, J., 2015. Preclinical Diagnosis of Magnetic Resonance (MR) Brain Images via Discrete Wavelet Packet Transform with Tsallis Entropy and Generalized Eigenvalue Proximal Support Vector Machine (GEPSVM). Entropy. 17(4), 1795-1813. https://doi.org/10.3390/e17041795.

Zhang, Z., 2017. A thermodynamics-based theory for the thermo-poro-mechanical modeling of saturated clay. International Journal of Plasticity. 92, 164-185. https://doi.org/10.1016/j.ijplas.2017.03.007.